\documentclass[debig,overfull]{epl}
\usepackage{graphicx}

\title{Casimir force between eccentric cylinders}

\author{Diego A.R. Dalvit \inst{1} \and Fernando C. Lombardo \inst{2} \and Francisco D. Mazzitelli \inst{2} \and
Roberto Onofrio \inst{3,4}}

\institute{
	\inst{1} Theoretical Division, MS B213, Los Alamos National Laboratory, Los Alamos, NM 87545, USA \\
	\inst{2} Departamento de F\'\i sica J.J. Giambiagi, Facultad de Ciencias Exactas y Naturales, 
	Universidad de Buenos Aires Ciudad Universitaria, Pabell\' on I, 1428 Buenos Aires, Argentina \\
	\inst{3} Department of Physics and Astronomy, Dartmouth College, 6127 Wilder Laboratory, Hanover, NH 03755, USA \\
	\inst{4} Dipartimento di Fisica ``G. Galilei'', Universit\`a di Padova, Via Marzolo 8, Padova 35131, Italy
}


\begin{document}

\maketitle

\begin{abstract}
We consider the Casimir interaction between a cylinder and a hollow 
cylinder, both conducting, with parallel axis and slightly different radii.  
The Casimir force, which vanishes in the coaxial situation, is evaluated 
for both small and large eccentricities using the proximity approximation. 
The cylindrical configuration offers various experimental advantages 
with respect to the parallel planes or the plane-sphere geometries, 
leading to favourable conditions for the search of extra-gravitational 
forces in the micrometer range and for the observation of finite temperature 
corrections.
\end{abstract}

Casimir forces are one of the most striking macroscopic
manifestations of vacuum quantum fluctuations.
Recently there has been an increasing interest in
experimental and theoretical aspects of these forces \cite{reviews}.
The force between parallel conducting plates as originally predicted in \cite{casimir}, 
after a first evidence reported in \cite{sparnaay}, has been recently measured at the 15\% 
accuracy level using cantilevers \cite{bressi}. 
A force of similar nature between a conducting plane and a conducting 
sphere, after pioneering studies \cite{vanblokland}, has also been 
investigated with progressively higher precision and accuracy. 
The latter force has been measured by using torsion balances \cite{lamoreaux}, 
atomic force microscopes \cite{mohideen}, and micromechanical resonators 
whose motion was detected through capacitance bridges \cite{chan1} 
and fiber optic interferometers \cite{decca}. 
Casimir forces may be relevant in nanotechnology, giving rise to 
interesting non-linear dynamics for nanoelectromechanical systems \cite{chan2}. 
Also, the predicted existence of new interactions with coupling comparable 
to gravity but range in the micrometer region \cite{fishbach} adds
strong motivations to control the Casimir force at the highest level of accuracy. 
This requires taking into account many deviations from the ideal situation 
initially discussed in \cite{casimir}, among these the corrections 
due to roughness and finite conductivity of the metallic surfaces, 
and the effect of the finite temperature, somewhat controversial 
\cite{mostep,hoye}, yet to be observed. Moreover, the observation 
of the thermal contribution is important in itself as a 
macroscopic test of quantum field theories at finite temperature.

In this paper we analyze a geometry different from the previously 
studied cases of two parallel planes or a plane-sphere.  
We consider one conducting cylinder inserted inside a hollow conducting 
cylinder, with parallel axis. In the ideal coaxial case the two 
axis will coincide and, based on symmetry arguments, this will result in a null 
Casimir force, while in the general eccentric case a finite force will arise. 
This configuration may be useful to minimize spurious gravitational
and electromagnetic effects, and has specific advantages with respect
to the parallel plane and the plane-sphere geometries.  
The discussion will naturally lead to some proposal for experimental 
schemes which should allow to get more stringent limits to extra-gravitational
forces in the micrometer range or to achieve 
an easier observation of the finite 
temperature corrections to the Casimir force.
 
Let us consider two eccentric cylinders of length $L$, with radii
$a$ and $b$ respectively (with $L \gg a, b$
to neglect border effects), as depicted in Fig. 1.
We will mainly focus on the particular case $a\simeq b$, since in 
this case the Casimir force is enhanced.  
The distance between the axis of the cylinders, a measure of the
eccentricity, will be denoted by $\epsilon$. 
In order to evaluate the Casimir energy we will use the proximity 
approximation \cite{derjaguin}. 
This is partially justified by recent results \cite{mazzitelli}
showing that, for concentric cylinders, the proximity approximation 
reproduces the exact results far beyond its expected range of validity. 
This result holds as long as one uses the geometric mean
prescription for the effective area \cite{gies}, a prescription which arises 
naturally in a semiclassical framework \cite{schaden,mazzitelli}.
It is then reasonable to assume that this is also true at 
least for slightly eccentric cylinders (the proximity force approximation
could be improved using the geometric optics approach put forward in \cite{jaffe},
or the numerical method of \cite{gies}). 

From the Casimir energy per unit area for parallel plates separated by a
distance $l$, 
\begin{equation}
E^{(0)}_{pp}(l)= -\frac{\pi^2 \hbar c}{720 l^3}\,\,\,,  \label{pp}
\end{equation}
the interaction energy between cylinders is, using the proximity approximation
\begin{equation}
E^{(0)}_I \simeq \int_0^{2\pi} dA_{\rm eff}(\theta ) E_{pp}^{(0)}(r(\theta) - a), 
\label{prox}
\end{equation}
where  $r(\theta )$ is the distance of a point of the external
cylinder to the axis of the inner one, and $dA_{\rm eff}(\theta)$
is the geometric mean of two infinitesimal adjacent areas on both cylinders. 
Following the notations introduced in Fig. 1 we find that $r(\theta )=
\sqrt{b^2 - \epsilon^2 \cos^2\theta} + \epsilon \sin\theta$ and
$dA_{\rm eff}= L\sqrt{ab+\epsilon a\sin\theta}d\theta$. 
The interaction energy and the force between the two cylinders 
$F^{(0)}_y=-\partial E_I/\partial \epsilon$
depend on the dimensionless parameters 
$\epsilon/b$ and $\tilde\epsilon=\epsilon/(b-a)$. 
Since we are considering $a\simeq b$, we will always have 
$\epsilon/b \ll 1$ and $\tilde\epsilon\gg \epsilon/b$. 
Thus to  lowest order in $\epsilon/b$ we obtain
\begin{equation}
F^{(0)}_y  =  -\frac{\pi^2 \hbar c L a}{240 (b-a)^4}
\int_0^{2\pi} \frac{d\theta \sin\theta}{( 1 +\tilde\epsilon
\sin\theta )^4} \simeq  \tilde\epsilon
\frac{1 +\frac{{\tilde\epsilon}^2}{4}}{\left
(1 -\tilde\epsilon^2\right)^{\frac{7}{2}}} F_0 , \label{result1}
\end{equation}
where $F_0=- \pi^3 \hbar c L a / 60 (b-a)^4$ is twice the equivalent Casimir
force between two parallel plates with the same area of the two cylinders 
and spaced by a distance $b-a$. 
It is worth to note that the force always makes unstable 
the equilibrium position of ideal coaxial cylinders.

\begin{figure}[t] 
\centerline{\includegraphics[width=11cm]{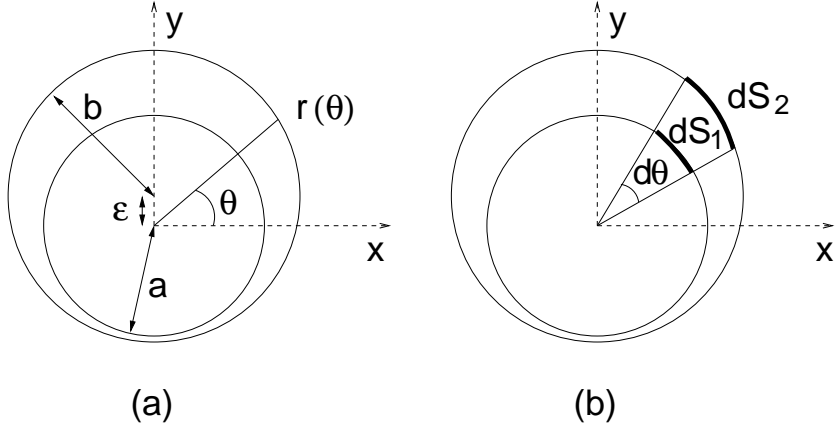}} 
\caption{Cylindrical geometry for measuring Casimir forces.  
(a) An inner cylinder of radius $a$ and a hollow cylinder of radius $b$, 
with the origin of coordinates on the axis of the inner cylinder, and  
distance $\epsilon$ between the two axis.
The function $r(\theta)$ gives the radial coordinate of the outer cylinder
from the axis of the inner one.
(b) The effective area for the application of the proximity
approximation, as the geometric mean of $dS_1$ and $dS_2$:  
$dA_{\rm eff}(\theta) =\sqrt{dS_1 dS_2}$.} 
\label{figure1}
\end{figure}
For nearly concentric cylinders, $\tilde\epsilon \ll 1$, the force
is linear in the distance between the axis of the cylinders
\begin{equation}
F^{(0)}_y \simeq  \frac{\epsilon}{b-a} F_0 .
\label{result2}
\end{equation}
This corresponds to an inverted harmonic oscillator, and explicitly shows
the instability. 
In the opposite case, when $\tilde\epsilon \rightarrow 1$, we get
\begin{equation}
F^{(0)}_y \simeq \frac{5}{32\sqrt 2} \left(\frac{b-a}{d}\right)^{7/2} F_0,
\label{result3}
\end{equation}
where $d=b-a-\epsilon$ is the distance between cylinders.
The behavior of the force in the large eccentricity limit 
$\tilde\epsilon \rightarrow 1$ is similar to that of a 
cylinder parallel to a plane. 
Indeed, with a cylinder of radius $a$ at a distance 
$d\ll a$ from a plane, the force is
\begin{equation}
F^{(0)}_{cp}(d) \simeq -{\frac{\pi^2\hbar c L}{120a^3}}
\int_0^{\pi/2}{\frac{d\theta}{(1+{d\over a}-\cos\theta)^4}} 
\simeq - {\pi^3\hbar c L a^{1/2}\over 384\sqrt 2 d^{7/2}}.
\label{cil-plane}
\end{equation}
The scaling of the Casimir force with distance is intermediate 
between the plane-spherical ($\propto d^{-3}$) and the parallel 
plate configuration ($\propto d^{-4}$), as well as the  absolute 
force signal, for typical values of the 
geometrical parameters. Indeed, 
the relative strengths of the forces for a common distance $d$ between 
the two bodies are:
\begin{eqnarray}
\frac{F_{pp}^{(0)}(d)}{F_{cp}^{(0)}(d)} &=&  
\frac{0.72 A}{L (a d)^{1/2}} ; \nonumber \\
\frac{F_{cp}^{(0)}(d)}{F_{sp}^{(0)}(d)} &=& 
\frac{0.66 L}{R} 
\left( \frac{a}{d} \right)^{1/2}, 
\end{eqnarray}
where $F_{pp}^{(0)}$ is the force between parallel plates of area $A$, 
$F_{sp}^{(0)}$ is the force between the sphere and the plane, with 
$R$ the radius of the sphere.
If, for instance, we choose $A = 1{\rm mm}^2$, $a=R=100\mu{\rm m}$, 
$L=5{\rm mm}$ and $d=1\mu{\rm m}$, the Casimir force ratios are 
$F_{pp}^{(0)}/F_{cp}^{(0)} \simeq 14$ and $F_{cp}^{(0)}/F_{sp}^{(0)} \simeq 330$.
With respect to the plane-spherical situation, one can enhance 
the signal by exploiting the linear dimension, {\it i.e.} the size $L$, 
at least as far as the parallelism between the axis of cylinder and
plane or their surface roughness do not become an issue. 

For an accurate comparison between experiment and theory we need 
to consider the deviations of the predicted force from the ideal 
situation of perfect conductors, zero temperature, and zero roughness.
For typical surfaces and realistic experimental sensitivities, 
roughness corrections are not relevant at the distances we are 
interested in ($d > 1 \mu{\rm m}$). On the other hand, combined
temperature and conductivity corrections are important in this
range of distances. These corrections have been computed using 
different approaches, leading to controversial predictions for
the Casimir force between parallel plates \cite{mostep,hoye,klimchitskaya}.

We have computed the combined corrections due to finite temperature and 
finite conductivity using two distinct theoretical models. 
In Figure 2a we show the combined corrections
obtained using the plasma model. Our starting point is the expression for the
interaction energy per unit area in the plane-plane geometry at finite temperature
and finite conductivity, $E_{pp}$, which was derived in \cite{klimchitskaya}
using the plasma model for the dielectric function in the Lifshitz formula.
Figure 2b depicts the combined corrections using the model described in 
\cite{hoye}, in which the transverse electric zero mode does not contribute
at all to the Casimir force. Our calculation is based on Fig. 4 of \cite{hoye},
which shows the surface force density between parallel gold metallic plates. 
Such force density was computed using the experimental data for the
permittivity of gold as a function of frequency \cite{astrid}.
The results for the different geometries shown in both figures have been numerically 
obtained from the plane-plane configuration using the proximity force approximation.
In the particular case of slightly eccentric cylinders, the Casimir force
can be easily obtained from $E_{pp}$. Indeed, 
the interaction energy between cylinders is given by Eq. (\ref{prox}), where 
$E_{pp}^{(0)}$ should be replaced by $E_{pp}$. 
Expanding the right hand side of Eq. (\ref{prox}) in powers of $\tilde{\epsilon}$ 
one can derive the force between nearly concentric cylinders 
($\tilde{\epsilon} \ll 1$) in terms of the second derivative of the energy  
$E_{pp}$ evaluated at a separation $l=b-a$, namely
\begin{equation}
F_y \simeq - \epsilon \pi L a \left.{\frac{d^2E_{pp}}{dl^2}}\right\vert_{(b-a)} .
\end{equation}

\begin{figure}[t]
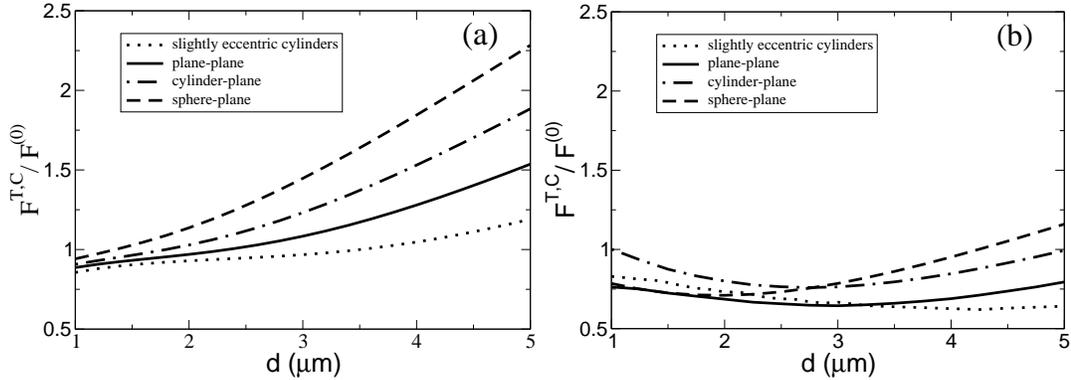
 
\includegraphics[width=7cm]{eccentric.fig2a.eps}
\includegraphics[width=7cm]{eccentric.fig2b.eps}
\caption{Combined thermal and conductivity corrections to the Casimir force for various geometries.
We depict the relative contributions $F^{T,C}/F^{(0)}$ versus the distance $d$ between gold metallic 
surfaces in the case of slightly eccentric cylinders, cylinder-plane,
plane-plane, and sphere-plane configurations, for two different finite
conductivity scenarios, the so-called plasma model (a) 
\cite{klimchitskaya} and a model that excludes the zero frequency TE
mode (b) \cite{hoye}.
For both figures parameters are $a=R= 100{\mu}{\rm m}$ and $T=300$ K.}
\label{figure3}
\end{figure}

We see that, in the range where temperature corrections 
can be more likely observable (above $\simeq 3 \mu$m) 
both models predict the same hierarchy for the various 
geometries, with the larger relative correction for the 
sphere-plane, followed by the cylinder-plane, plane-plane, 
and slightly eccentric cylinders, respectively. 
At the same time, the corrections in the same range of distances 
are significantly different to allow for a crucial test of the models, 
with an enhancement and a depletion of the measurable force with 
respect to the zero-temperature case for the two models respectively, 
resulting in predicted forces differing by almost a factor 2.  
The absolute magnitude of the force in the plane-cylindrical case is 
much larger than that of the sphere-plane situation and, with respect 
to the parallel plates, there are less issues of dust and parallelism, 
making this configuration more favourable for looking at thermal corrections.
Instead, in the case of slightly eccentric cylinders the corrections 
are weaker than those in the other three geometries, making this 
configuration more robust for seeking extra-gravitational forces with
suppressed background.

We now discuss possible experimental arrangements for measuring 
the Casimir force between cylinders. 
In the case of the almost coaxial configuration $\tilde\epsilon\ll 1$, 
one possibility is to repeat a microscopic version of the experiment 
described in \cite{hoskins} to test universal gravitation in the cm range, 
with a small torsional balance mounted on the ends of the internal cylinder. 
In this case the unstable force could be evidenced by intentionally creating 
a controlled eccentricity and measuring the feedback force required to 
bring the internal cylinder to zero eccentricity, as depicted in Fig. 3a.  
A somewhat simpler situation can be imagined by attaching the external hollow 
cylinder to a cantilever, then creating a resonator of effective 
mass $M$ and natural angular frequency $\omega_0$, see Fig. 3b. 
In the presence of the inner cylinder (kept in a fixed position), 
the frequency of the resonator for small oscillations around the 
equilibrium position is renormalized by the negative spring 
constant of the Casimir force (see Eq. (\ref{result2})). 
Assuming a small frequency shift ({\it i.e.} a Casimir force much 
smaller than the restoring force of the resonator), and 
achievable values $a=100 {\mu{\rm m}}$, $L=5 {\rm mm}$, $M=10^{-6} 
{\rm kg}$, $b-a=1 \mu$m and $\omega_0=10^3 {\rm s^{-1}}$, 
we obtain 
$\Delta \omega/\omega_0=- F_0 / 2(b-a)M \omega_0^2 =
-4.25 \times 10^{-3}$, which is
within the sensitivity of frequency-shift techniques 
on microresonators \cite{bressi2,decca}. 

\begin{figure}[t] 
\centerline{\includegraphics[width=11cm]{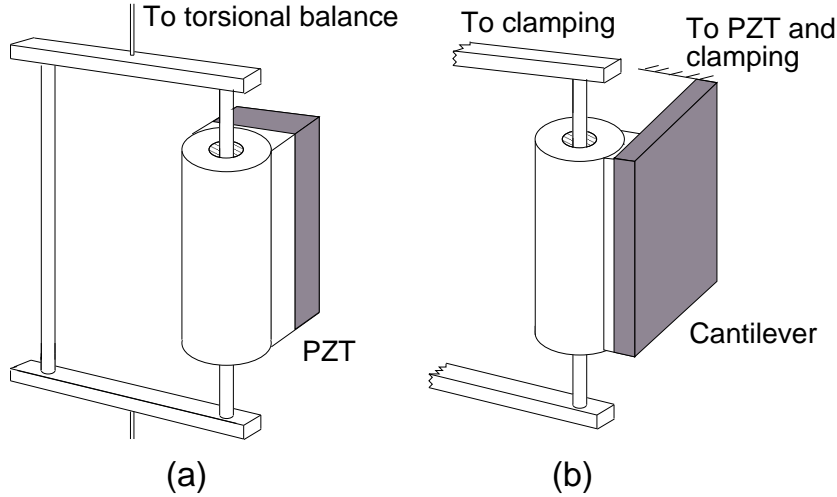}} 
\caption{Experimental schemes for detecting Casimir forces with 
slightly eccentric cylinders. 
(a) The inner cylinder is rigidly connected to a torsional balance 
and the signal to restore the zero eccentricity configuration after a
controlled displacement is monitored.  
(b) The hollow cylinder is connected to a cantilever and the frequency 
shift induced in the small oscillations is measured.}
\label{figure2}
\end{figure}

This configuration has some advantages over the parallel plates geometry. 
If there is no residual charge in the inner cylinder, the system remains 
neutral and screened by the external one from background
noise sources, and from residual charges in the outer cylinder. 
When the inner cylinder has a residual charge, there will be a 
small potential difference $V$ between the cylinders, and the 
coaxial configuration will be electrostatically unstable. 
The Laplace equation for the electrostatic 
potential $\phi$ in the region between the two cylinders can be solved by imposing 
the boundary conditions $\phi(a,\theta)=0$ and
$\phi(r(\theta),\theta)=V$. 
To first order in $\epsilon/b$ we find
\begin{equation}
\phi(r,\theta) \simeq \frac{V}{\log(b/a)}
\left[
\log \left( \frac{r}{a} \right) -
\epsilon \frac{r^2-a^2}{b^2-a^2} \frac{\sin\theta}{r}
\right] .
\label{phi}
\end{equation}
The electrostatic force between cylinders can be computed as
$F_y^E \simeq \epsilon_0 \pi V^2 L a \epsilon/(b-a)^3$, 
where $\epsilon_0$ is the electric
permittivity of vacuum. This result, which shows the electrostatic
instability, can also be obtained from $F_y^E= -(\partial
U_E/\partial \epsilon)_V$, using a proximity approximation for the
electrostatic energy $U_E$
\begin{equation}
U_E \simeq \frac{1}{2} \epsilon_0 V^2 L a  \int_0^{2\pi} \frac{d\theta}{r(\theta ) - a} .
\label{elec}
\end{equation}
The electrostatic instability can be avoided by putting the 
cylinders in contact, something unavoidable during the preliminary 
stages of parallelization. 
Then the residual charge of the inner cylinder will flow to the hollow cylinder, 
apart from a residual charge due to imperfections and finite length of the cylinders.
This residual charge will be smaller than for other geometries, as the same 
discharging procedure does not work in the other configurations. 
If some residual potential difference 
still remains, it will contribute to the frequency shift, however  
it can be eliminated by a counterbias as in all the  
Casimir experiments performed so far.
The electrostatic instability can be exploited to improve the parallelism between cylinders. 
One could apply a time-dependent potential between the cylinders 
and measure the force, as in the experiments to test the inverse-square Coulomb law. 
Parallelism and concentricity would be maximum for a minimum value of the force.
Moreover, the expected gravitational force is obviously null, this being an advantage to 
look for intrinsically short-range extra-gravitational forces.
These should violate the Gauss law, as it could be evidenced by performing an experiment analogous to that done
with macroscopic cylinders in \cite{hoskins}. 

In conclusion, we have computed the Casimir force between conducting  
eccentric cylinders, using the proximity approximation, also
including finite temperature and conductivity effects.  
Our results suggest that cylindrical configurations could be useful 
to precisely measure the Casimir force and related signals superimposed
to it. 
We have briefly described experimental configurations which look promising, 
either to minimize spurious effects of gravitational, electrostatic, conductivity
and thermal origin to search for new forces, as in the case of slightly eccentric cylinders, or 
for intentionally looking at finite-temperature corrections to the Casimir force,
as in the cylindrical-plane configuration.
Recent progress in machining nanomechanical structures 
should make our proposal feasible.

\acknowledgments
We would like to thank A. Roncaglia and P. Villar for computational help. 
This work  was supported by DOE, University of Buenos Aires, CONICET,
Fundaci\'on Antorchas and ANPCyT.


\begin{thebibliography}{99}

\bibitem{reviews} 
PLUNIEN G., M\"ULLER B. and GREINER W., Phys. Rep., {\bf 134} (1986) 87;
MILONNI P., {\it The quantum vacuum}, (Academic Press, San Diego, 1994);
MOSTEPANENKO V. M. and TRUNOV N. N., {\it The Casimir effect and its
  applications}, (Clarendon, London, 1997); 
BORDAG M., {\it The Casimir effect 50 years later}, (World Scientific,
Singapore, 1999);  
BORDAG M.,  MOHIDEEN U., and MOSTEPANENKO V. M., 
Phys. Rep., {\bf 353} (2001) 1;
REYNAUD S. {\it et al.}, C. R. Acad. Sci. Paris {\bf IV-2} (2001) 1287;
MILTON K. A., {\it The Casimir Effect: Physical Manifestations of the 
Zero-Point Energy} (World Scientific, Singapore, 2001).

\bibitem{casimir} CASIMIR H. B. G., Proc. K. Ned. Akad. Wet. B, {\bf 51}
(1948) 793.

\bibitem{sparnaay} SPARNAAY M. J., Physica, {\bf 24} (1958) 751.

\bibitem{bressi} BRESSI G., CARUGNO G., ONOFRIO R. and RUOSO G.,
Phys. Rev. Lett., {\bf 88} (2002) 041804.

\bibitem{vanblokland} VAN BLOKLAND P. H. G. M. and OVERBEEK J. T. G., 
J. Chem. Soc. Faraday Trans. I, {\bf 74} (1978) 2637.

\bibitem{lamoreaux} LAMOREAUX S. K., Phys. Rev. Lett., {\bf 78} (1997) 5.

\bibitem{mohideen} MOHIDEEN U. and ROY A., Phys. Rev. Lett., {\bf 81} (1998) 4549;
HARRIS B. W., CHEN F., and MOHIDEEN U., Phys. Rev. A, {\bf 62} (2000) 052109.

\bibitem{chan1} CHAN H. B., AKSYUK V.A., KLEIMAN R.N., BISHOP D.J. and CAPASSO F.,
Science, {\bf 291} (2001) 1941.

\bibitem{decca} DECCA R. S., LOPEZ D., FISCHBACH E. and KRAUSE D.E.,
Phys. Rev. Lett., {\bf 91} (2003) 050402.

\bibitem{chan2} CHAN H. B., AKSYUK V.A., KLEIMAN R.N., BISHOP D.J. and CAPASSO F.,
Phys. Rev. Lett., {\bf 87} (2001) 211801.

\bibitem{fishbach} FISCHBACH E. and TALMADGE C. L., 
{\it The search for Non-Newtonian gravity} (AIP/Springer-Verlag, New York, 1999).

\bibitem{mostep} 
BOSTROM M. and SERNELIUS B. E., Phys. Rev. Lett., {\bf 84} (2000) 4757; 
LAMOREAUX S. K., Phys. Rev. Lett., {\bf 87} (2001) 139101; 
SERNELIUS B. E., Phys. Rev. Lett., {\bf 87} (2001) 139102;
CHEN F., KLIMCHITSKAYA G. L., MOHIDEEN U., and MOSTEPANENKO V. M., 
Phys. Rev. Lett., {\bf 90} (2003) 160404; 
GEYER B., KLIMCHITSKAYA G. L., and MOSTEPANENKO V. M., Phys. Rev. A, {\bf 67} (2003) 062102; 
{\it ibid.} {\bf 65} (2002) 062109; 
ESQUIVEL R., VILLARREAL C., and MOCHAN W. L., quant-ph/0306139;  
MOCHAN W. L., VILLARREAL C., and ESQUIVEL-SIRVENT R., quant-ph/0206119; 
GENET C., LAMBRECHT A., and REYNAUD S., Int. J. Mod.  Phys. A, {\bf 17} (2002) 761; 
Phys. Rev. A, {\bf 62} (2000) 012110; 
SVETOVOY V. B. and LOKHANIN M. V., Phys. Rev. A, {\bf 67} (2003) 022113.

\bibitem{hoye}
H{\O}YE J. S., BREVIK I., ASRSETH J.B. and MILTON K.A.,
Phys. Rev. E, {\bf 67} (2003) 056116.

\bibitem{derjaguin} DERJAGUIN B. V. and ABRIKOSOVA I. I., Sov. Phys. JETP,
{\bf 3} (1957) 819; DERJAGUIN B. V., Sci. Am., {\bf 203} (1960) 47.

\bibitem{mazzitelli} MAZZITELLI F. D., SANCHEZ M.J., SCOCCOLA N.N. and VON STECHER J.,
Phys. Rev. A, {\bf 67} (2003) 013807.

\bibitem{gies} Similar results based on numerical simulations have 
been reported for the force between a plate and a sphere, see:
GIES H., LANGFELD K., and MOYAERTS L., J. High Energy Phys., {\bf 6} (2003) 018.

\bibitem{schaden} SCHADEN M. and SPRUCH L., Phys. Rev. Lett., {\bf 84} (2000) 459; 
Phys. Rev. A, {\bf 58} (1998) 935.

\bibitem{jaffe} JAFFE R. L., SCARDICCHIO A., Phys. Rev. Lett., {\bf 92} (2004) 070402.

\bibitem{klimchitskaya} KLIMCHITSKAYA G. L. and MOSTEPANENKO V. M., 
Phys. Rev. A, {\bf 63} (2001) 062108.

\bibitem{astrid} LAMBRECHT A., REYNAUD S., Eur. Phys. J. D {\bf 8} (2000) 309;
Phys. Rev. Lett. {\bf 84} (2000) 5672.

\bibitem{hoskins} HOSKINS J. K., NEWMAN R.D., SPERO R. and SCHULTZ J., 
Phys. Rev. D, {\bf 32} (1985) 3084.

\bibitem{bressi2} BRESSI G., CARUGNO G., GALVANI A., ONOFRIO R., RUOSO G. and VERONESE F.,
Class. Quantum Grav., {\bf 18} (2001) 3943. 



\end{thebibliography}
\end{document}